\documentclass[english,aps,prl, twocolumn,showpacs]{revtex4}
\pdfoutput=1
\usepackage[latin1]{inputenc}
\usepackage{amsmath}
\usepackage{graphicx}
\usepackage{amssymb}

\makeatletter

\@ifundefined{textcolor}{}
{%
 \definecolor{BLACK}{gray}{0}
 \definecolor{WHITE}{gray}{1}
 \definecolor{RED}{rgb}{1,0,0}
 \definecolor{GREEN}{rgb}{0,1,0}
 \definecolor{BLUE}{rgb}{0,0,1}
 \definecolor{CYAN}{cmyk}{1,0,0,0}
 \definecolor{MAGENTA}{cmyk}{0,1,0,0}
 \definecolor{YELLOW}{cmyk}{0,0,1,0}
 }

\makeatother

\usepackage{babel}

\begin{document}

\title{AC-Conductance through an Interacting Quantum Dot}

\author{Björn Kubala and Florian Marquardt}

\affiliation{Physics Department, Arnold Sommerfeld Center for Theoretical Physics,
and Center for NanoScience, Ludwig-Maximilians-Universität, 80333
Munich, Germany}
\begin{abstract}
We investigate the linear ac-conductance for tunneling through an
arbitrary interacting quantum dot in the presence of a finite dc-bias.
In analogy to the well-known Meir-Wingreen formula for the dc case,
we are able to derive a general formula for the ac-conductance. It
can be expressed entirely in terms of local correlations on the quantum
dot, in the form of a Keldysh block diagram with four external legs.
We illustrate the use of this formula as a starting point for diagrammatic
calculations by considering the ac-conductance of the noninteracting
resonant level model and deriving the result for the lowest order
of electron-phonon coupling. We show how known results are recovered
in the appropriate limits.
\end{abstract}

\pacs{71.38.-k, 72.10.Bg, 72.30.+q, 73.23.-b}

\maketitle

\paragraph{Introduction.-}

Recent experimental progress enables the study of the dynamics of
electronic processes in mesoscopic structures. In a seminal experiment,
Gabelli et al.~\citep{Gabelli_Science06} measured the ac-conductance
of a mesoscopic RC-circuit. Measuring the in and out of phase components
of the ac-conductance at GHz frequency, they confirmed theoretical
predictions for the coherent transport of noninteracting electrons
\citep{Buttiker1993Dynamic,Buttiker1993PLA,Pretre1996Dynamic}. General
interest in time- or frequency-dependent properties of mesoscopic
systems is driven by efforts to measure and manipulate quantum coherent
systems as quickly as possible, for instance, for applications in
quantum information transmission and processing. An important building
block for such experiments may be provided by a coherent single-electron
source recently realized \citep{Feve2007OnDemand}. Furthermore, time-
or frequency-dependent measurements can yield additional information
about internal time and energy scales that are not accessible via
the time-averaged dc-current. 

A cornerstone for many recent theoretical considerations of the time-independent
dc-current was laid in the 1992 paper of Meir and Wingreen \citep{MeirPRL92},
who derived a Landauer-type formula, giving the current through a
central interacting region in terms of local Keldysh Green's functions
of the central region (and the tunnel-coupling and Fermi-functions
of the leads). In this paper we extend this approach to the finite
frequency case and derive an expression for the current response to
a small ac-voltage excitation, which is applied across the central
interacting region in addition to a static dc-bias, i.e., the linear
ac-conductance in a nonequilibrium situation. As in the Meir-Wingreen
approach we derive a formula for the current in terms of `local' objects,
but in our case these are no longer the usual Green's functions, i.e.,
expectation values of two-operator objects, but contain four electronic
operators. The actual evaluation of these central objects may in principle
be performed employing various techniques and approximations. The
approach lends itself, however, to a combination with a diagrammatic
calculation perturbative in the interaction strength \citep{Rammer1986Quantum},
as we demonstrate in an examplary calculation for electron-phonon
coupling.

Before turning to the general approach, we note some previous works
that have considered special cases. For vanishing electron-phonon
coupling our approach reproduces calculations based on the scattering
matrix method \citep{Buttiker1993Dynamic,Pretre1996Dynamic,Fu1993Quantum}.
Those calculations rely on a noninteracting single-particle picture,
but can be amended to capture screening by incorporating the effects
of induced charges on the internal potentials in a self-consistent
manner. The current response to a fixed internal potential as calculated
by the scattering method (or by our approach) is then used as an input
for these calculations. Another route to finite frequency calculations
starts from an expression of the current in Green's function formulation
for completely general time dependence of the parameters \citep{JauhoPRB94}.
Progress can then again be made for the noninteracting case \citep{JauhoPRB94},
or employing special approximations to the interacting self-energy
and Green's function \citep{Li1996Quantum}. Kubo relations naturally
link the linear ac-conductance to current-current correlators. Indeed,
as pointed out in \citep{Gavish_2001} its real part is related to
the asymmetric part of the frequency dependent current-noise for a
static dc-bias. Derivations of a general Meir-Wingreen type formula
for the noise, however, have been restricted to the symmetrized noise
and, furthermore, consider the noninteracting case \citep{Bo1996Lowfrequency}
or focus on the zero frequency limit \citep{Zhu_PRB03,Galperin_PRB06}.

\paragraph{System.-}

\begin{figure}
\includegraphics[clip,width=1\columnwidth]{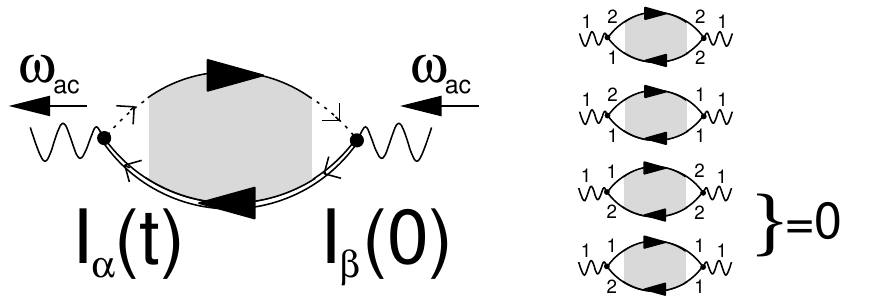} 

\caption{\label{fig:indexing}Correlator needed for the ac conductance. One
lead-electron line (dotted line) and one dot-electron line (double
line) are connected to each outer vertex. The two contributions in
the current vertex, Eq.~\eqref{eq:current vertices}, corresponding
to an in- or outgoing dot-electron line (and vice versa for the lead
electron) yield different signs for the various configurations. The
four diagrams at the right show the possible combinations of Keldysh
indices at the inner legs, with the lower two vanishing. A single
solid line stands for either dot-electron or lead-electron lines.}

\end{figure}

We consider transport through a central region, for instance a quantum
dot, between a left and right lead. While electrons in the leads can
be considered as noninteracting, within the central region there might
be electron-electron or (as considered in the specific calculations
in this paper) electron-phonon interaction. The Hamiltonian of the
system can then be written as $\hat{H}=\hat{H}_{\text{central}}+\sum_{\alpha=L/R}(\hat{H}_{\textrm{res},\alpha}+\hat{H}_{T,\alpha})\,$.
Lead Hamiltonian $\hat{H}_{\text{res,}\alpha}=\sum_{k}\epsilon_{k\alpha}\hat{c}_{k\alpha}^{\dagger}\hat{c}_{k\alpha}$
and tunnel Hamiltonian $\hat{H}_{T,\alpha}=\sum_{k,\, n}V_{k\alpha,n}\hat{c}_{k\alpha}^{\dagger}\hat{d}_{n}+\, c.c.\,$
$ $are of the standard form and the central part, $\hat{H}_{\textrm{central}}=\sum_{n}\varepsilon_{n}\hat{d}_{n}^{\dagger}\hat{d}_{n}+\hat{H}_{\textrm{int}}$
may include electron-electron or electron-phonon coupling within $\hat{H}_{\textrm{int}}$.
We want to consider the nonequilibrium situation, where on top of
a finite dc-bias a small ac-excitation voltage $V_{\beta}$ of frequency
$\omega_{\textrm{ac}}$ is applied to lead $\beta$. The resulting
finite frequency current $I_{\alpha}$ flowing from lead $\alpha$
to the central region is given by the linear ac-conductance, $I_{\alpha}(\omega_{\textrm{ac}})=\mathfrak{\mathcal{G}}_{\alpha\beta}(\omega_{\textrm{ac}})\, V_{\beta}(\omega_{\textrm{ac}})\,.$
Considering linear response with respect to the ac-excitation, we
can use Kubo formalism, but due to the finite dc-bias the expectation
values in the Kubo-formalism have to be taken for the nonequilibrium
finite-bias state. The standard Kubo approach directly yields the
linear conductance $\mathcal{G}_{\alpha\beta}$ in terms of a (retarded)
current-current correlator, $\mathcal{G}_{\alpha\beta}(\omega_{\textrm{ac}})=\left[K_{\alpha\beta}(\omega_{\textrm{ac}})-K_{\alpha\beta}(0)\right]/(i\omega_{\textrm{ac}})\,$,
where\begin{equation}
K_{\alpha\beta}(\omega_{\textrm{ac}})=-\frac{i}{\hbar}\int_{0}^{\infty}dt\; e^{i\omega_{\textrm{ac}}t}\langle\left[\hat{I}_{\alpha}(t),\;\hat{I}_{\beta}(0)\right]\rangle\,.\label{eq:correlator_def}\end{equation}
The current operators are of similar form as the tunnel Hamiltonian,
\begin{eqnarray}
\!\!\!\!-\frac{1}{(-e)}\hat{I}_{\alpha}(t) & = & \dot{\hat{N}}_{\alpha}(t)=\frac{i}{\hbar}[\hat{H},\,\hat{N}_{\alpha}]=\frac{i}{\hbar}[\hat{H}_{T,\alpha},\,\hat{N}_{\alpha}]\nonumber \\
 & = & -\frac{i}{\hbar}\sum_{kn}(V_{k\alpha,n}\hat{c}_{k\alpha}^{\dagger}\hat{d}_{n}-V_{n,k\alpha}\hat{d}_{n}^{\dagger}\hat{c}_{k\alpha})\,.\label{eq:current vertices}\end{eqnarray}
The correlator $K_{\alpha\beta}(t,t')=K_{\alpha\beta}(t-t',0)$ has
the same structure as a retarded Green's function, but with a pair
of creation and destruction operators for each of the current operators
at the two time variables, see the schematic diagrammatic representation
in Fig.~\ref{fig:indexing}. Each current operator appears as an
external vertex with one lead-electron line (dotted line) and one
dot-electron line (double line), one of them entering, one leaving
the vertex %
\footnote{Note that the dot-electron lines include the effects of tunneling
excursions to the leads.%
}. The two possibilities (for each vertex) carry signs according to
Eq.~\eqref{eq:current vertices}. In general, in between the four
legs starting from the two vertices there will be a complicated diagram
involving many interaction lines, as indicated by the shaded box in
Fig.~\ref{fig:indexing}. The simplest type of diagram, however,
is encountered in the non-interacting case. Then all operators are
directly contracted, resulting in two Green's function running antiparallel
(one starts at $t$ going to $t'$, the other goes from $t'$ to $t$).
Langreth's theorem then yields so-called analytical continuation rules,
giving the proper combination of nonequilibrium Green's function for
the different Keldysh components of the product \citep{Rammer1986Quantum}.
For instance, the object with two antiparallel Green's functions $C(t,t')=A(t,t')B(t',t)$
has the retarded component of $C^{R}(t,t')=A^{<}(t,t')B^{A}(t',t)+A^{R}(t,t')B^{<}(t,t')=\frac{1}{2}(A^{K}B^{A}+A^{R}B^{K})$.
These rules correspond to applying the standard rules for Keldysh
indices at the outer vertices. (We use the following matrix representation
of the nonequilibrium Green's functions $\left(\begin{array}{cc}
G^{R} & G^{K}\\
0 & G^{A}\end{array}\right)=\left(\begin{array}{cc}
G^{11} & G^{12}\\
G^{21} & G^{22}\end{array}\right)$, where the Keldysh Green's function component $G^{21}$ vanishes
by construction.) Turning to the interacting case, we use these same
rules for the external vertices of the full correlator. We find four
possible combination of Keldysh indices on the inner legs of the correlator
loop, see Fig.~\ref{fig:indexing}. Note that two of these combinations
would include the vanishing Green's function component $G^{21}$ without
further interaction lines added to the simple loop diagram. In that
case, the Langreth result is directly recovered. In the following
we will argue, that the two latter index combinations vanish also
after including all interaction lines and we are left with only the
upper two combinations in Fig.~\ref{fig:indexing}. %

To start with, we consider only one additional interaction line. Then
it is easy to see that this statement holds, simply by constructing
all diagrams according to the indexing rules at the vertices. Vanishing
diagrams are then found for two reasons. First, the application of
the Keldysh indexing rules at the vertices may lead to the appearance
of a line corresponding to the vanishing Green's function component
$G^{21}$. Secondly, diagrams may also vanish due to the fact that
the appearance of an advanced or retarded Green's function between
two vertices implies a fixed order of the time variables of the two
interconnected vertices. Such a time order can lead to a logical contradiction,
if, for instance, the Green's functions implied that $t_{1}<t_{2}<t_{3}$
and $t_{3}<t_{1}$. 

\begin{figure}
\includegraphics[width=1\columnwidth]{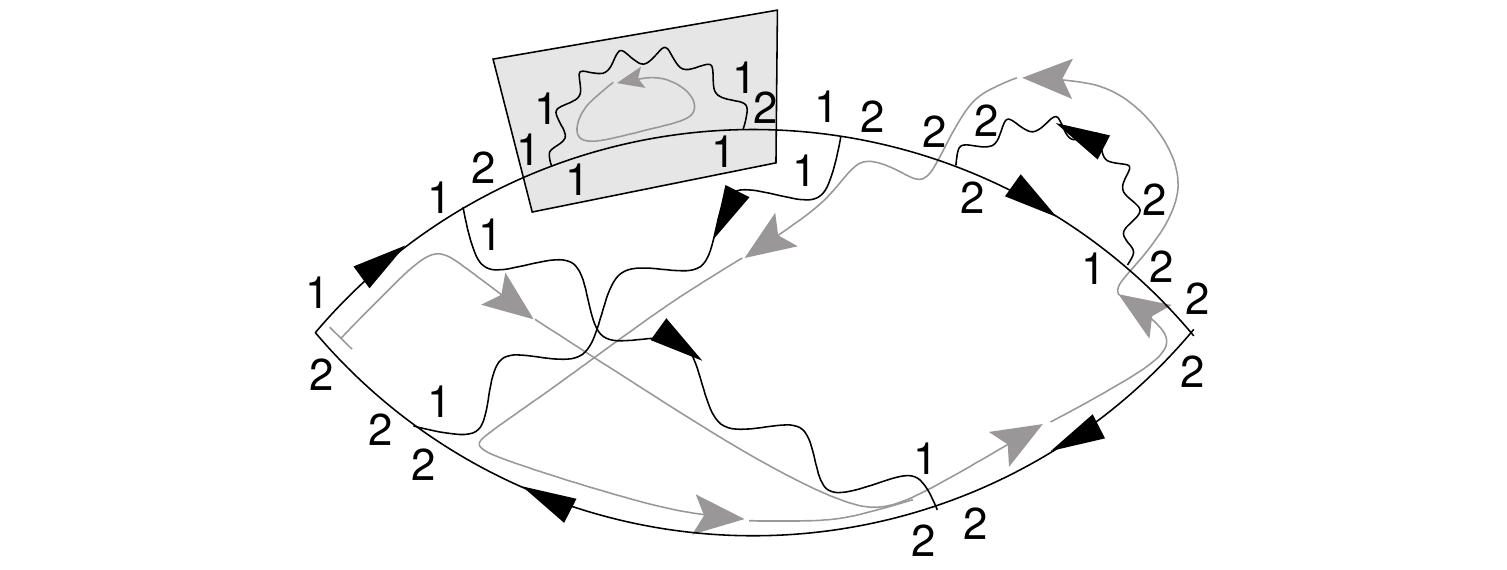}

\caption{\label{fig:time-arrow lines forbid diagrams}Time-arrow lines, running
parallel to $G^{11}$ or antiparallel to $G^{22}$, do not stop internally,
but progress through any possible internal vertex. They may hit themselves,
forming a closed loop, which indicates a logical contradiction in
the time ordering, or end at the external vertices. Additional time-arrow
lines (cf.~the shaded box) may also exist and enforce vanishing of
a diagram if they form a loop.}

\end{figure}

To exploit this property systematically for all higher order diagrams,
we introduce the notion of a \emph{time-arrow line}, which appears
in diagrams indicating a fixed time-ordering of the vertices along
this line, see Fig.~\ref{fig:time-arrow lines forbid diagrams}.
Such a line runs parallel to $G^{11}=G^{R}$ and antiparallel to $G^{22}=G^{A}$.
Analysing the 12 different possible index combinations at a vertex
(we consider the electron-phonon case here, but the argument is general
and relies on the Keldysh indexing rules at the vertices only) we
find that if a time arrow-line runs into a vertex, there will always
also be such a line leaving the vertex. In consequence a time-arrow
line will never stop at an internal vertex, but will run through a
diagram until it hits an outer vertex (where it may end or go on)
or until it hits itself and forms a closed loop. Such a closed loop
will imply a logical contradiction within the time ordering as discussed
above, and the corresponding diagram vanishes. Consider now a combination
of Keldysh indices at the outer vertices as in the diagram shown in
Fig.~\ref{fig:time-arrow lines forbid diagrams}. Due to the vanishing
of $G^{21}$ we can infer that two time-arrow lines start at the left
external vertex, while at the right vertex we can have a time-arrow
line starting or crossing through the vertex. As the time-arrow lines
do not stop internally and can not stop at either of the two external
vertices in this specific diagram, they have to form a closed loop.
Analysing the external vertices in this manner, we can thus state
that some index combinations will necessarily result in time-arrow
loops and are, hence, forbidden. For the case of interest here, the
two lower diagrams of Fig.~\ref{fig:indexing} are of this type and
vanish.%
\begin{figure}[b]
\includegraphics[width=1\columnwidth]{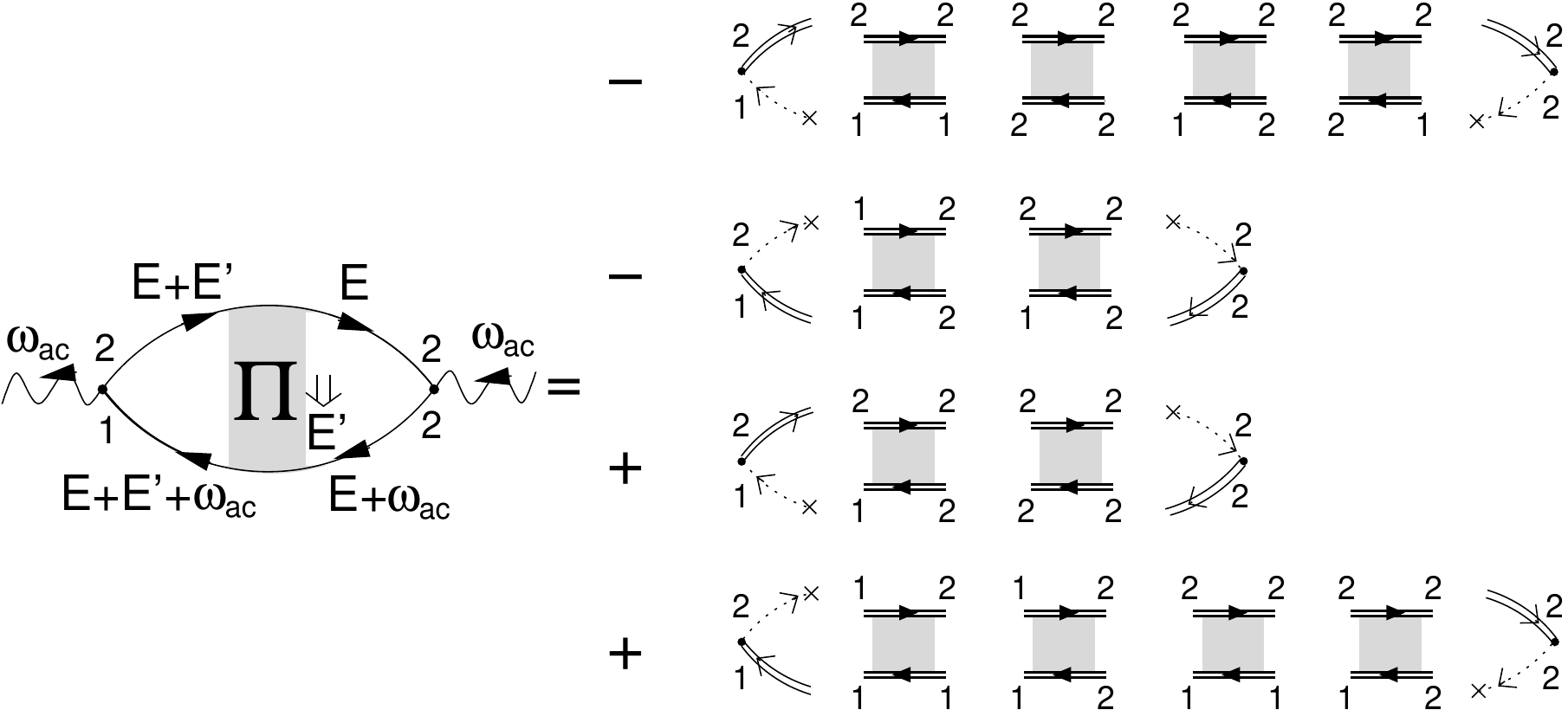}

\caption{\label{fig:splitting off lead Gs}Each correlator contains four combinations
of in-/out-tunneling at the outer vertices. The dot-lead Green's functions
are split into lead-lead and dot-dot Green's functions, so that the
correlator can be expressed in terms of a central object $\Pi$ containing
dot operators only, represented by gray boxes (with Keldysh indices
at the legs indicated). For example, $_{1}^{2}\Pi_{1}^{2}$ occurs
in the first and last line, and these diagrams contribute to the term
displayed in Eq.~(\ref{eq:correlator in terms of dot-propagators}).}

\end{figure}

\paragraph{Separating out the leads.-}

Each of the two external vertices is connected to a lead-electron
line. Moving away from the vertices along these lines, we will encounter
a tunneling event changing the lead-electron line to a dot-electron
line %
\footnote{As a special case, we have to account for the possibility of directly
connecting the lead electron lines from the two external vertices
without the inclusion of a dot-electron Green's function. This extra
`direct' contribution is only possible, then the two vertices involve
lead-electron operators from the same lead, i.e., for the correlator
$K_{\alpha\alpha}$.%
}. Note, that there can be no interaction line in between the external
vertex and the first tunneling, as interaction is only present within
the dot region. The Green's function corresponding to the considered
line is now split into a product of a lead-lead Green's function,
a tunneling matrix element and a dot-electron Green's function. The
Keldysh indexing of those Green's functions follows the rules of matrix
multiplication in Keldysh space, see Fig.~\ref{fig:splitting off lead Gs}.
This split-off procedure has to be performed for the four possible
orientations of dot and lead lines at the external vertices (taking
account of the proper signs according to Eq.~\eqref{eq:current vertices}).
We proceed by evaluating the sums over lead states contained in the
outer current vertices and in the inner tunneling vertices marked
by crosses in Fig.~\ref{fig:splitting off lead Gs}. For that, we
change from time to frequency/energy domain. We introduce the frequency
$E$, which encircles the loop, and the energy $E'$ exchanged within
the central object between upper and lower line (see Fig.~\ref{fig:splitting off lead Gs}).
The summations are then executed under the standard assumption of
the wide-band limit to yield $\sum_{k}|V_{\alpha}|^{2}g_{\alpha}^{R/A}=\mp\frac{i}{2}\Gamma_{\alpha}\,,\,\,\sum_{k}|V_{\alpha}|^{2}g_{\alpha}^{K}=-i\Gamma_{\alpha}(1-2f^{\alpha})\,,$
where $g_{\alpha}^{R/A/K}$ are the bare lead Green's functions, $f^{\alpha}$
the Fermi-distribution function of lead $\alpha$ and $\Gamma_{\alpha}$
is the coupling strength defined as $\Gamma_{\alpha}=2\pi\sum_{k}|V_{\alpha}|^{2}$.
For ease of notation we consider a single level in the central region
here and in the following. The simple extension to the general case
is discussed below. 

Separating out the leads in that manner we are left with a central
object $\Pi$ with four legs of dot-electron lines, which includes
all possible further interaction lines in its central block, see Fig.~\ref{fig:splitting off lead Gs}.
Starting from the two remaining diagrams of Fig.~\ref{fig:indexing},
we can group together all contributions with the same indices on the
central object $\Pi$. We find that all but three contributions cancel,
and the final result for the frequency-dependent retarded current-current
correlator can be written as\begin{eqnarray}
K_{\alpha\beta}^{C} & = & -i\frac{e^{2}}{\hbar}\Gamma_{\alpha}\Gamma_{\beta}\int\frac{dE}{2\pi}\int\frac{dE'}{2\pi}\left[f_{E}^{\beta}-f_{E+\omega_{\textrm{ac}}}^{\beta}\right]\times\label{eq:correlator in terms of dot-propagators}\\
 & \times & \!\!\!\left[_{1}^{2}\Pi_{1}^{2}+(1-2f_{E+E'+\omega_{\textrm{ac}}}^{\alpha}){}_{2}^{2}\Pi_{1}^{2}-(1-2f_{E+E'}^{\alpha}){}_{1}^{1}\Pi_{1}^{2}\right]\nonumber \end{eqnarray}
plus extra terms for the diagonal conductances, stemming from a direct
contraction of lead operators of the outer vertices, $K_{\alpha\alpha}^{\textrm{direct}}=-\frac{e^{2}}{\hbar}\Gamma_{\alpha}\int\frac{dE}{2\pi}\,[f_{E+\omega_{\textrm{ac}}}^{\alpha}G_{E}^{A}+f_{E-\omega_{\textrm{ac}}}^{\alpha}G_{E}^{R}]\,,$
so that $K_{\alpha\beta}=K_{\alpha\beta}^{C}+\delta_{\alpha\beta}K_{\alpha\alpha}^{\textrm{direct}}\,.$

This constitutes a general expression for the ac-conductance in terms
of a central `dot-operator only' object, which is the equivalent to
the dot Green's functions in the Meir-Wingreen result. This central
object $\Pi$ is a Keldysh-type diagram with four external legs. Two
dot-electron lines enter the diagram and two leave it. We arrange
the diagram so that the lines entering are placed at the upper left
and lower right corner. Due to energy conservation the energies carried
by the four external lines can be parametrized by three parameters
$E,E',$ and $\omega_{\textrm{ac}}$ as indicated in Fig.~\ref{fig:splitting off lead Gs}.
The four external legs carry Keldysh indices as indicated on $\Pi$.
The central object $\Pi$ then contains all possible diagrams within
these specifications, fully including tunneling and interaction.

If we wish to consider several levels within the central dot-region,
each dot-line carries extra level indices, in general differing at
start and end (due to tunneling to the leads and back into a different
level). The couplings $\Gamma_{\alpha}$ become matrices in these
dot indices and the trace is taken over the whole expression. For
instance, the first term of Eq.~\eqref{eq:correlator in terms of dot-propagators}
would change into $\Gamma_{\alpha,\, ij}\,\Gamma_{\beta,\, kl}\;_{j1}^{i2}\Pi_{1l}^{2k}$
with summation over all level indices $i,j,k,l$. In the following,
we illustrate the application of our formula for two simple examples.

\paragraph{Resonant level model.-}

Without interaction, there are no phononic lines connecting the upper
and lower line of $\Pi$ and energy exchange between these lines is
not possible. The superfluous $E'$ integral is then cancelled by
a $\delta$-function guaranteeing energy conservation. The terms in
Eq.~\eqref{eq:correlator in terms of dot-propagators} with $_{2}^{2}\Pi_{1}^{2}$
and $_{1}^{1}\Pi_{1}^{2}$ vanish (since they would contain $G^{21}=0$),
while the $_{1}^{2}\Pi_{1}^{2}$ contribution and the direct part
immediately yield\begin{eqnarray*}
\mathcal{G}_{\alpha\beta}^{(0)} & = & \frac{e^{2}}{\hbar}\frac{\Gamma_{\alpha}\Gamma_{\beta}}{\omega_{\textrm{ac}}}\int\frac{dE}{2\pi}(f_{E+\omega_{\textrm{ac}}}^{\beta}-f_{E}^{\beta})G_{E}^{A,\, T}G_{E+\omega_{\textrm{ac}}}^{R,\, T}\\
 &  & \!\!\!\!\!\!\!\!+\delta_{\alpha\beta}\frac{e^{2}}{\hbar}\frac{i\Gamma_{\alpha}}{\omega_{\textrm{ac}}}\int\frac{dE}{2\pi}(f_{E+\omega_{\textrm{ac}}}^{\beta}\!\!-f_{E}^{\beta})(G_{E}^{A,\, T}-G_{E+\omega_{\textrm{ac}}}^{R,\, T})\,.\end{eqnarray*}
The noninteracting dot Green's functions including the tunneling self-energy
are $G^{R/A,\, T}(E)=[E-\varepsilon\pm i(\Gamma_{L}+\Gamma_{R})/2]^{-1}\,$.
We thus recover the result that can also be derived using methods
specialized to the non-interacting case, viz. the fully time-dependent
Green's function approach of Ref. \citep{JauhoPRB94} or scattering
matrix theory \citep{Pretre1996Dynamic}.

\paragraph{Second-order in e-ph coupling.-}

\begin{figure}[h]
\includegraphics[width=1\columnwidth]{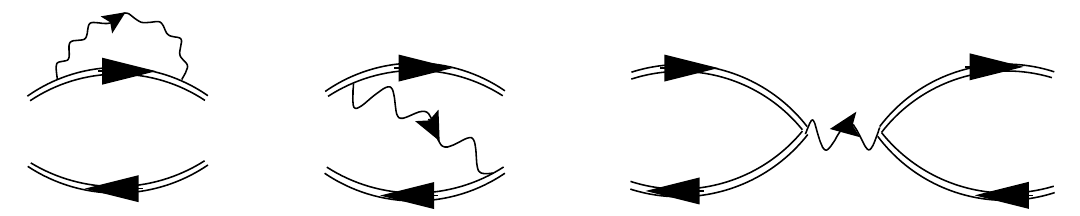}

\caption{\label{fig:Two-types-of 2nd order}Three types of second-order diagrams.
Diagrams with one Green's function dressed with a phonon line (left),
diagrams with one phonon line crossing the loop (middle), and Hartree-type
diagrams of two electronic loops coupled by a phononic line (right).}

\end{figure}

We now consider coupling to phononic modes of frequency $\omega_{\kappa}$,
i.e. we take $\hat{H}_{\textrm{int}}=\sum_{\kappa}[g_{\kappa}(\hat{a}_{\kappa}+\hat{a}_{\kappa}^{\dagger})\hat{d}^{\dagger}\hat{d}+\hbar\omega_{\kappa}\hat{a}_{\kappa}^{\dagger}\hat{a}_{\kappa}]\,$.
We find that in second order in the e-ph coupling $g_{\kappa}$, i.e.,
with one phonon line included in the diagram, there are three different
type of contributions. Firstly , we can replace one of the bare dot-electron
Green's function by one dressed with a phonon line, see Fig.~\ref{fig:Two-types-of 2nd order}.
For the evaluation of this type of diagrams, we accordingly replace
bare by dressed Green's function in the zeroth-order result and expand
to second order, e.g., $G_{E}^{A,\, T}\rightarrow G_{E}^{A,\, T}\Sigma_{E}^{A,\,\textrm{ph}}G_{E}^{A,\, T}\,$,
where the self-energies $\Sigma^{\textrm{ph}}$ are the lowest order
Fock- and Hartree-terms, which depend on the bare phonon propagator
$D_{\kappa}^{R/A/K}\,.$ Secondly, there are contributions with one
crossing phonon line, which yield\begin{eqnarray*}
\!\! K_{\alpha\beta}^{(2),\,\textrm{cross}} & \!\!\!\!\!\!=\!\!\!\!\! & -i\frac{e^{2}}{\hbar}\Gamma_{\alpha}\Gamma_{\beta}\int\frac{dE}{2\pi}\int\frac{dE'}{2\pi}\left[f_{E}^{\beta}-f_{E+\omega_{\textrm{ac}}}^{\beta}\right]\times\\
 &  & \,\!\!\!\!\!\!\!\!\!\!\!\!\!\!\!\!\!\!\times i\sum_{\kappa}\frac{g_{\kappa}^{2}}{2}G_{E}^{A,\, T}G_{E+\omega_{\textrm{ac}}}^{R,\, T}\left\{ \left[G_{E+E'}^{A,\, T}D_{\kappa,\, E'}^{A}G_{E+E'+\omega_{\textrm{ac}}}^{K,\, T}\right.\right.\\
 &  & \!\!\!\!\!\!\!\!\!\!\!\!\!\!\!\!\!\!\!\!\!\!\left.+\, G_{E+E'}^{K,\, T}D_{\kappa,\, E'}^{R}G_{E+E'+\omega_{\textrm{ac}}}^{R,\, T}\!\!+G_{E+E'}^{A,\, T}D_{\kappa,\, E'}^{K}G_{E+E'+\omega_{\textrm{ac}}}^{R,\, T}\right]\\
 &  & \!+\left[(1-2f_{E+E'+\omega_{\textrm{ac}}}^{\alpha})G_{E+E'}^{A,\, T}D_{\kappa,\, E'}^{A}G_{E+E'+\omega_{\textrm{ac}}}^{A,\, T}\right]\\
 &  & \!-\left[(1-2f_{E+E'}^{\alpha})G_{E+E'}^{R,\, T}D_{\kappa,\, E'}^{R}G_{E+E'+\omega_{\textrm{ac}}}^{R,\, T}\right]\,,\end{eqnarray*}
where the three terms in the square brackets correspond to the three
different indexing variations of $\Pi$ in Eq.~\eqref{eq:correlator in terms of dot-propagators}.
The contribution from the third type of diagram in Fig.~\ref{fig:Two-types-of 2nd order}
vanishes in the dc-limit in linear response and will not be discussed
further here. %

The equivalence of the Meir-Wingreen result to the dc-limit of our
result is most easily seen for the special case of linear response.
Then the explicit difference of Fermi functions $f^{\beta}$ in the
last two equations results in a derivative, while all other Fermi
and Green's functions can be taken at equilibrium and vanishing $\omega_{\textrm{ac}}\,$.
Contributions from diagrams with one crossing phonon line can then
be written\[
\mathcal{G}_{\alpha\beta}^{(2),\textrm{cross}}\!=\! i\frac{e^{2}}{\hbar}\frac{\Gamma_{\alpha}\Gamma_{\beta}}{\Gamma_{L}+\Gamma_{R}}\!\!\int\!\!\frac{dE}{2\pi}f_{E}^{\beta\prime}G_{E}^{R,\, T}\!\!\left[\Sigma_{E}^{R,\,\textrm{ph}}\!\!-\Sigma_{E}^{A,\,\textrm{ph}}\!\right]\! G_{E}^{A}.\]
Expanding the Green's function in the Meir-Wingreen result to second
order in the electron-phonon coupling we finally recover the proper
dc-limit. A slightly more involved calculation confirms the equivalence
for finite dc-bias.

\paragraph{Conclusions.-}

We derived a formula for the linear ac-conductance, Eq.~\eqref{eq:correlator in terms of dot-propagators},
which is applicable for arbitrary interactions and for a finite applied
dc-bias. The formula expresses the ac-conductance in terms of a central
object $\Pi$, which can be calculated using Keldysh diagrammatic
methods. We illustrated the application of our formula by explicit
calculations for the resonant level model and the case of electron-phonon
coupling.

We acknowledge funding by the DFG via SFB/TR 12, SFB 631, NIM, and
the Emmy-Noether program and thank A.~Clerk for a critical reading of the manuscript.

\bibliographystyle{apsrev}
\bibliography{ac-conductance_shortpaper}

\end{document}